\documentclass[12pt,preprint]{aastex}
\usepackage{graphicx}
\usepackage{amsmath}
\usepackage{subfigure}

\newcommand{\kms}{\rm ~km~s^{-1}}

\begin{document}
\title{TIME-DEPENDENT DIFFUSIVE SHOCK ACCELERATION IN SLOW SUPERNOVA REMNANT SHOCKS}
\author{Xiaping Tang and Roger A. Chevalier}
\affil{Department of Astronomy, University of Virginia, P.O. Box 400325, \\
Charlottesville, VA 22904-4325; xt5ur@virginia.edu, rac5x@virginia.edu}

\begin{abstract}
Recent gamma ray observations show that middle aged supernova remnants interacting with molecular clouds can be sources of both GeV and TeV emission. Models involving  re-acceleration of pre-existing cosmic rays in the ambient medium and direct interaction between supernova remnant and molecular clouds have been proposed to explain the observed gamma ray emission. For the re-acceleration process, standard diffusive shock acceleration theory in the test particle limit produces a steady state particle spectrum that is too flat compared to observations, which suggests that the high energy part of the observed spectrum has not yet reached a steady state.   We derive a time dependent DSA solution in the test particle limit for situations involving re-acceleration of pre-existing cosmic rays in the preshock medium. Simple estimates with our time dependent DSA solution plus a molecular cloud interaction model can reproduce the overall shape of the spectra of IC 443 and W44  from GeV to TeV energies through pure $\pi^0$-decay emission.
We allow for a power law momentum dependence of the diffusion coefficient, finding that a power law  index of 0.5 is favored.

\end{abstract}

\keywords{acceleration of particles --- gamma-rays: ISM --- ISM: individual objects (IC 443,W44) --- ISM: supernova remnants}

\section{INTRODUCTION}
Diffusive shock acceleration (DSA) is believed to be the particle acceleration mechanism in most astrophysical environments involving shock waves \citep[e.g.,][]{B&E87}.
The theory naturally produces a power law energy spectrum of energetic particles in the steady state.
Accelerated  particles can produce $\gamma$-ray emission through either bremsstrahlung and inverse Compton emission of leptonic origin, or $\pi^0$-decay emission of hadronic origin, making $\gamma$-ray observations  important diagnostics for particle acceleration processes in astronomical objects. Recent observations from both space-based GeV observatories and ground-based TeV observatories show that middle aged supernova remnants (SNRs) interacting with molecular clouds can be sources of both GeV \citep{uchiyama11} and TeV \citep{Albert07,Aharonian08,Aleksic12} emission. The characteristic $\pi^0$-decay signature identified in IC 443 and W44 \citep{Giuliani11,Ackermann13} provides possible direct evidence for cosmic ray (CR) particle acceleration in supernova remnants.  

Two scenarios have been proposed to explain the observed GeV and TeV emission from  middle aged SNRs. 
In one, nearby molecular clumps are illuminated  by the accelerated CR particles escaping from a SNR in addition to the pre-existing CR background,  producing the GeV and TeV emission \citep{Gabici09,Fujita09,LC10,Ohira11};  the other  involves direct interaction between the SNR and the molecular clumps \citep{bykov00,Uchiyama10,Inoue10,Tang14}. 
In \cite{Tang14}  we noted evidence that in middle aged remnants with both GeV and TeV emission (IC 443, W28, W51C), the emission regions are co-located and spatially correlated with the shocked molecular clump region \citep{Abdo10,uchiyama11,Nicholas12}, which indicates there is direct interaction between the SNR and molecular clumps.
In the direct interaction scenario, re-acceleration of pre-existing CRs has been considered while particle injection through the thermal pool is neglected in view of the slow radiative shock \citep{B&C82,Uchiyama10}. Recent observations of nearby giant molecular clouds by {\it Fermi} reveal $\gamma$-ray emission as a result of  interaction between the Galactic CR background and  giant molecular clouds \citep{Yang14}, showing the importance of the pre-existing CR component. The standard DSA theory produces too flat a steady state particle spectrum compared to that indicated by observations, so it has been suggested that there is insufficient time to reach the steady state particle spectrum in the energy region of interest; the upper limit based on the acceleration timescale has been implemented as an
exponential cutoff in the particle spectrum \citep{Uchiyama10,Tang14}.
While this model compares well to data in the GeV range, it falls below the observations in
the TeV range \citep{Uchiyama10,Tang14}.
\cite{Tang14} further found that a model in which the energetic particles are compressed in
the radiative shock fronts with no DSA is able to reproduce the observed spectral shapes of
the high energy $\gamma$-ray emission.
Although this model has attractive features, it requires a high covering factor for the
shock wave emission and it is unclear why DSA is not occurring.

Here, we examine in more detail the case where DSA is occurring, but it has not had time
to reach a steady state.
Time dependent test particle DSA was first discussed  in detail by \cite{Toptygin80}.
An analytic solution for continuous injection of a monoenergetic spectrum at the shock front with source  term $S=\delta(x)\delta(p-p_0)$, where $p$ is particle momentum and $p_0$ is the injected momentum,  was obtained in the special case that the shock velocity $U$ and diffusion coefficient $\kappa$ are constant, and the ratio $\kappa_1/U_1^2=\kappa_2/U_2^2$, where the subscript 1 refers to upstream and 2 to downstream. \cite{Forman83}  solved the case that $\kappa$ has a power law dependence on momentum in the limit that $p\gg p_0$.
\cite{Drury91} extended the solution to a more general case in which the flow velocity $U$ and diffusion coefficient $\kappa$ also have spatial dependence.  

The monoenergetic spectrum assumption made in the above discussions made it possible to decouple the time dependent solution into the product of the steady state solution and a time evolution factor, which is useful for investigating the acceleration timescale for individual particle in the system.  The resulting acceleration time is in good agreement with the discussion from the microscopic method \citep{Lagage83}. However there has been less attention  to the evolution of the spectral shape for a group of particles with an arbitrary spectrum. 
Here, we consider the case where the upstream region is filled  with seed particles, in particular, pre-existing CRs. 
We  limit our discussion to DSA in the test particle limit for simplicity, as has been assumed in
previous discussions of middle aged remnants with slow shock waves \citep{B&C82,Uchiyama10}

 In Section 2, the time dependent DSA solution for a shock wave interacting with pre-existing CRs is derived for both energy independent diffusion and energy dependent diffusion. We then calculate the $\pi^0$-decay emission from IC 443 and W44 based on our time dependent solution in Section 3 and compare the results to observations.  Aspects of our model are discussed in Section 4.

\section{PARTICLE SPECTRUM}

We consider a plane parallel shock wave and constrain our discussion to the shock frame.
The shock front is at $x=0$ and the flow is moving toward the positive $x$ direction with flow velocity $U=U_1+(U_2-U_1)H(x)$, where the subscripts 1 and  2 refer to upstream and downstream, respectively, throughout the paper and $H(x)$ is the step function. 
Here, we are mostly interested in the radiative phase of the SNR when the shock is slow and the test particle theory may be applicable. In the early phases of the SNR, the shock is fast and  non-linear effects may be important. As a result, the accelerated particle distribution in the early phase of a SNR is  difficult to model.  Fortunately, the total number of CR particles accelerated in a remnant before the radiative phase is likely to be small compared to the pre-existing CRs  swept up in the radiative phase for the energy range we are interested in.  The remnant spends most of its time and sweeps up most of its volume in the radiative phase.  We ignore the particles  accelerated in the  early phase of the remnant to simplify the calculation and  consider only re-acceleration of pre-existing CRs in the radiative phase. We investigate a situation with seed particles in the shock upstream region; the shock front starts to interact with the seed particles at time $t=0$. The advection diffusion equation we need to solve  becomes \citep{Drury83}
\begin{equation}
\frac{\partial f}{\partial t} +U\frac{\partial f}{\partial x}=\frac{\partial}{\partial x}\left(\kappa\frac{\partial f}{\partial x }\right)+\frac{\partial U}{\partial x} \left( \frac{p\; \partial f}{3\;  \partial p}\right) +Q(p)\delta (t)H(-x),
\label{ad-diff}
\end{equation}
where $f(x,t,p)$ is the isotropic part of the particle phase space density, $\kappa=\kappa_1+(\kappa_2-\kappa_1)H(x)$ is the diffusion coefficient and $Q(p)\delta (t)H(-x)$ is the source term representing the pre-existing CRs in the  upstream region. The above equation is  correct only when $v\gg U \gg U_A$, where $v$ is the particle velocity and $U_A$ is the Alfven velocity for the magnetic irregularities. The condition $v\gg U$ implies that the result cannot be applied to relativistic shocks. The condition $U\gg U_A$ requires that the magnetic field cannot be too strong or  second order Fermi acceleration would be important and we would need to add a momentum diffusion term to  equation (\ref{ad-diff}). The presence of the shock discontinuity requires a matching condition at the shock front to solve the equation. The easiest way to obtain the matching condition is to integrate equation (\ref{ad-diff}) from $x=-0$ to $x=+0$ with weight function 1 and $\int dx/\kappa$. The resulting matching conditions are
\begin{equation}
\left[ f \right]^{+0}_{-0}=0 \qquad \mbox{  and  }   \qquad
\left[\kappa\frac{\partial f}{\partial x}+\frac{U}{3}p\frac{\partial f}{\partial p}\right]^{+0}_{-0}=0.  
\end{equation}
 
In this paper, we are primarily interested at the time evolution of the spectral shape in the downstream region, which is determined by the ratio $t/\tau$,  where $\tau$ is a time scale characterizing the DSA in the system that will be chosen in a form to simplify the calculation. It is more convenient to use the dimensionless time factor $t/\tau$ than time $t$.  Defining $\theta=t/\tau$ and assuming $h(x,\theta,p)=f(x,t,p)$, we find that the advection  diffusion equation  becomes
\begin{equation}
\frac{1}{\tau}\frac{\partial h}{ \partial \theta} +U\frac{\partial h}{\partial x}=\frac{\partial}{\partial x}\left(\kappa\frac{\partial h}{\partial x }\right)+Q(p)H(-x)\frac{\delta (\theta)}{\tau},
\label{diff-h}
\end{equation}
while the matching conditions are now 
 \begin{equation}
\left[ h \right]^{+0}_{-0}=0  \qquad  \mbox{  and  }   \qquad
\left[\kappa\frac{\partial h}{\partial x}+\frac{U}{3}p\frac{\partial h}{\partial p}\right]^{+0}_{-0}=0.
 \label{match-cond}   
\end{equation}

Following the procedure in \cite{Drury91}, we perform a Laplace transform of the advection diffusion equation, assuming 
\begin{equation}
g(x,s,p)=\int^\infty_0 h(x,\theta,p)e^{-s\theta}d\theta.
\end{equation}
Then equation (\ref{diff-h}) becomes 
\begin{equation}
\frac{sg}{\tau}+U\frac{\partial g}{\partial x}=\kappa\frac{\partial^2 g}{\partial x^2}+\frac{Q(p)H(-x)}{\tau}.
\end{equation}
Since $f\rightarrow 0$ when $x\rightarrow +\infty$  and $f\rightarrow Q(p)H(t)H(-x)$ when $x \rightarrow -\infty$, we have $g\rightarrow 0$ when $x\rightarrow +\infty$  and $g\rightarrow Q(p)H(-x)/s$ when $x \rightarrow -\infty$. With these boundary conditions, the solution to the ordinary differential equation has the form 
 \begin{equation}
 g(x,s,p)=\begin{cases} c_1(s,p)e^{(A_1+2)xU_1/2\kappa_1}+Q(p)/s, & \mbox{if } x<0\\
 c_2(s,p)e^{-A_2xU_2/2\kappa_2},&\mbox{if } x>0, \end{cases}
 \label{Inverse-f}
 \end{equation}
where 
\begin{equation}
A_1=\sqrt{1+\tau_1s/ \tau}-1,   \qquad  A_2=\sqrt{1+\tau_2s/ \tau}-1,
\label{param}
\end{equation}
and
\begin{equation}
\tau_1=\frac{4\kappa_1}{U_1^2},   \qquad  \tau_2=\frac{4\kappa_2}{U_2^2}.
\end{equation}
The quantities $c_1(s,p)$ and $c_2(s,p)$ can be calculated by applying the matching conditions,  equation (\ref{match-cond}). After some calculation we obtain the downstream solution 
\begin{eqnarray}
\begin{aligned}
g(x,s,p)&=c_2(s,p)e^{-A_2xU_2/2\kappa_2}\\
&=\alpha p^{-\alpha}e^{-A_2xU_2/2\kappa_2}\int_0^p p'^{\alpha-1}\, Q(p')\, dp' \,\frac{A_1+2}{2s}\,e^{ -\int^p_{p'}\Delta \alpha \,  dp''/p''}
\end{aligned}
\label{postshock}
\end{eqnarray}
where $\alpha=3U_1/(U_1-U_2)$ and $\Delta \alpha=3(U_1A_1+U_2A_2)/2(U_1-U_2)$. 

In this paper we limit our discussion to the situation that the shock velocity $U$ is constant  while the diffusion coefficient $\kappa$ can have an energy dependence. We assume $\kappa$ is constant in space because the spatial dependence of $\kappa$ requires detailed information about the shock structure, which is beyond the discussion here. The spatially independent diffusion coefficient $\kappa$ applied in the following discussion can be considered as a spatially averaged value. We assume $\tau_1=\tau_2$ for simplicity, which requires $\kappa_1=16\kappa_2$ for a strong shock by the definition $\tau=4\kappa/U^2$. It is a strong constraint not always satisfied in actual situations.   We denote the quantities in the general situation with a 
hat symbol ($\hat{}$). In the case $\hat{\tau_1}\neq\hat{\tau_2}$, the spectral shape of the time dependent solution is mainly determined by the $\max\left\lbrace\hat{\tau_1},\hat{\tau_2}\right\rbrace$ since $\hat{\tau_1}$ and $\hat{\tau_2}$ together characterize the DSA time scale of the system and the simplified $\tau$ derived in our spectral fits can be considered as a good approximation to the $\max\left\lbrace\hat{\tau_1},\hat{\tau_2}\right\rbrace$, i.e. $\tau=\tau_1=\tau_2\approx \max\left\lbrace\hat{\tau_1},\hat{\tau_2}\right\rbrace$. For the SNR forward shock, it is likely that $\hat{\kappa_1}\gg16\hat{\kappa_2}$ and $\hat{\tau_1}\gg \hat{\tau_2}$, because the diffusion coefficient in the upstream region gradually increases from a value close to Bohm limit near the shock front to the value of the CR diffusion coefficient in the ISM far away from the shock front. The result $\tau \approx\max\left\lbrace\hat{\tau_1},\hat{\tau_2}\right\rbrace=\hat{\tau_1}$ thus can provide information on the spatially averaged diffusion coefficient in the upstream region, which can also be taken as an upper limit to the CR diffusion coefficient close to the shock front.  We start with the case where $\kappa$ is independent of energy and then discuss the situation where $\kappa$ has a power law dependence on particle momentum.

\subsection{Energy Independent Diffusion}  

For the case $U$ and $\kappa$ are constant and satisfy $\tau_1=\tau_2$, the analytic solution for $f(x,t,p)$ can be obtained by performing an inverse Laplace transform on equation (\ref{postshock}). Defining $\tau=\tau_1=\tau_2$, we obtain the analytic solution 
\begin{eqnarray}
\begin{aligned}
f(x,t,p)&=h(x,\theta,p)=\mathcal{L}^{-1}\left\lbrace g(x,s,p) \right\rbrace\\
&=\alpha p^{-\alpha}\int_0^p p'^{\alpha-1}\, Q(p')\, dp'\, \frac{1}{2} \left\{ \frac{1}{\sqrt{\pi \theta}}e^{-w^2}+{\rm erfc}(w)\right\},
\end{aligned}
\label{TD_solution}
\end{eqnarray}
where $w=\beta/2 \sqrt{\theta}-\sqrt{\theta}$ and
$\beta=\Delta \alpha {\rm ln}(p/p')/A_2+xU_2/2\kappa_2$.
The result can also be obtained by integrating the solution in \cite{Toptygin80} over the source position $z_0$ from $-\infty$ to $0$.

\begin{figure}[htb]
\begin{center}
\includegraphics[width=\textwidth]{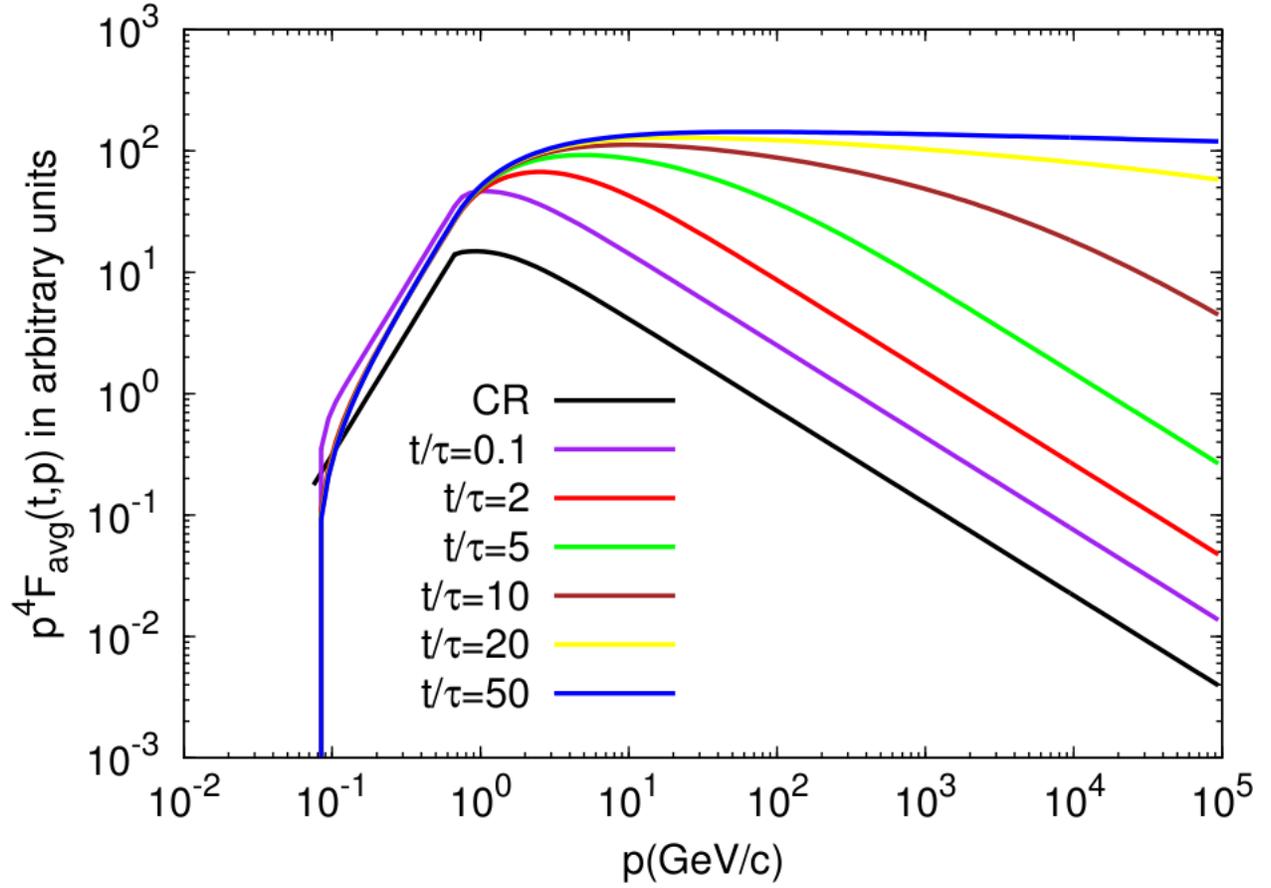} 
    
    \caption{Spatially averaged downstream particle momentum spectrum for various time ratios; see text for details. The cosmic ray spectrum denoted by CR has arbitrary scaling.}
    \label{DSA_integration}
 \end{center}
 \end{figure}

We assume strong shock conditions throughout the paper except for the slow molecular shock discussed later
and that $Q(p)$ follows the same CR spectrum as in \cite{Tang14}. The calculated downstream particle spectrum at the shock front $(x=0)$ for various $\theta$ is shown in Fig\ \ref{DSA_sf}. In the Appendix, we provide a simple argument for understanding the resulting spectral shape with energy independent diffusion.  There is a critical momentum below which the spectrum reaches the steady state solution, while above it the spectrum recovers the steep power law shape of the input CR spectrum at high energy.
The particle spectrum of interest for comparison with observations is the spectrum of all the accumulated particles integrated over the downstream region,
which is calculated by  
\begin{eqnarray}
\begin{aligned}
F(t,p)&=\mathcal{L}^{-1}\left\lbrace \int^\infty_0 g(x,s,p)dx \right\rbrace \\
&=\frac{U_2t}{2\theta}\alpha p^{-\alpha}\int_0^p p'^{\alpha-1} Q(p') dp' \left[ \sqrt{\frac{\theta}{ \pi}}  e^{-b^2}+\left(\frac{1}{2}-b\sqrt{\theta}\right){\rm erfc}(b) \right],
\end{aligned}
\label{TD_integrate}
\end{eqnarray}    
where $b=\Delta \alpha {\rm ln}(p/p')/2A_2\sqrt \theta-\sqrt{\theta}$. For a planar shock, $U_1t$ is the length scale of preshock medium swept up by the shock at time $t$, while $U_2t$ is the length scale of the shocked medium accumulated in the  postshock region. The quantity $F_{avg}(t,p)= F(t,p)/U_2t$ then can be considered as the spatially averaged downstream phase space density. We plot the spatially averaged downstream particle spectrum $F_{avg}(t,p)$ as a function of $p$ for various $\theta$ in Fig\ \ref{DSA_integration}. The spectral shape evolution of $F_{avg}(t,p)$ basically follows the same trend as for $f(0,t,p)$, with a critical energy characterizing the shape of the resulting spectrum. The transition between the steady state solution and the steep power law shape of the input CR spectrum is smoother for the spatially averaged case. The critical energy discussed here is different from the maximum energy defined in a situation in which a monoenergetic particle input spectrum is assumed. In that case there is a maximum energy that particles can achieve during the acceleration process. For our
situation of re-acceleration of pre-existing CR, the concept of
maximum energy is not relevant to the critical energy
discussed here.

\subsection{Energy Dependent Diffusion}
When $\kappa$  depends on particle momentum, the situation becomes more complicated. \cite{Forman83} provide the solution for a diffusion coefficient $\kappa$ with a power law energy dependence but for a monoenergetic input spectrum and particle momentum much greater than particle injected momentum ($p\gg p_0$). Thus we cannot use their solution to investigate the evolution of the spectral shape. Here we assume $\kappa=\overline{\kappa}\,\overline{p}^\sigma$, where $\overline{p}=pc/{(\rm 1~GeV)}$ is the dimensionless particle momentum, and $\overline{\tau_1}=\overline{\tau_2}=4\overline{\kappa_1}/U_1^2=4\overline{\kappa_2}/U_2^2$ to simplify the calculation. Following the procedure we used to solve the energy independent diffusion case, the spatially integrated particle spectrum in the downstream region is now  
\begin{eqnarray}
\begin{aligned}
F(t,p)&=\mathcal{L}^{-1}\left\lbrace \int^\infty_0 g(x,s,p)dx \right\rbrace \\
&=\mathcal{L}^{-1} \left\lbrace \alpha p^{-\alpha}\int_0^p p'^{\alpha-1}\, Q(p')\, dp' \,\frac{\kappa_2[A_1(p')+2]}{A_2(p)U_2s}\,e^{ -\int^p_{p'}\Delta \alpha \,  dp''/p''} \right\rbrace,
\end{aligned}
\label{postshock_ED}
\end{eqnarray}
where all the parameters are the same as defined before except we now have an energy dependent diffusion coefficient $\kappa=\overline{\kappa}\,\overline{p}^{\sigma}$.   Taking $\tau= \overline{\tau_1}=\overline{\tau_2} $,  we have
\begin{eqnarray}
\begin{aligned}
g(s,p)&=  \int^\infty_0 g(x,s,p)dx =\alpha p^{-\alpha} \int_0^p p'^{\alpha-1} Q(p')dp' \frac{\kappa_2 }{U_2}
\\ &\times\frac{e^{2\Delta \alpha(\sqrt{1+sp^\prime_\sigma }-\sqrt{1+sp_\sigma})/ A_2\sigma }(1+\sqrt{1+sp_\sigma})^{(2\Delta \alpha / A_2\sigma)+1}}{p_\sigma s^2(1+\sqrt{1+sp'_\sigma})^{(2\Delta \alpha / A_2\sigma)-1}},
\end{aligned}
\end{eqnarray}
where $p_\sigma=\overline{p}^\sigma$. The corresponding spatially averaged downstream particle spectrum is
\begin{eqnarray}
\begin{aligned}
F_{avg}(t,p)&=\frac{f_d(t,p)}{U_2t}=\frac{\alpha}{4 \theta} p^{-\alpha} \int_0^p p'^{\alpha-1} Q(p')dp' \\
&\times\mathcal{L}^{-1}\left\lbrace \frac{e^{2\Delta \alpha(\sqrt{1+sp'_\sigma }-\sqrt{1+sp_\sigma})/A_2\sigma} (1+\sqrt{1+sp_\sigma})^{(2\Delta \alpha / A_2\sigma)+1}}{s^2(1+\sqrt{1+sp'_\sigma})^{(2\Delta \alpha / A_2\sigma)-1}}  \right\rbrace. 
\end{aligned}
\label{ED_solution}
\end{eqnarray}
It is difficult to calculate the above inverse Laplace transform analytically, so we used Talbot's method to do the inversion of the Laplace transform numerically \citep{Talbot79}.
Before we applied Talbot's method to equation (\ref{ED_solution}), we did some tests of the numerical method by comparing the numerical results with the analytical solution we derived for the energy independent diffusion case.  The results based on Talbot's method were completely consistent with the analytical solution. We note that the energy dependent solution in equation (\ref{ED_solution}) cannot be extended to the energy independent diffusion case with $\sigma=0$. 

\begin{figure}[htb]
\begin{center} 
\subfigure[] 
{\label{DSA_ED_1}    
 \includegraphics[width=0.48\textwidth]{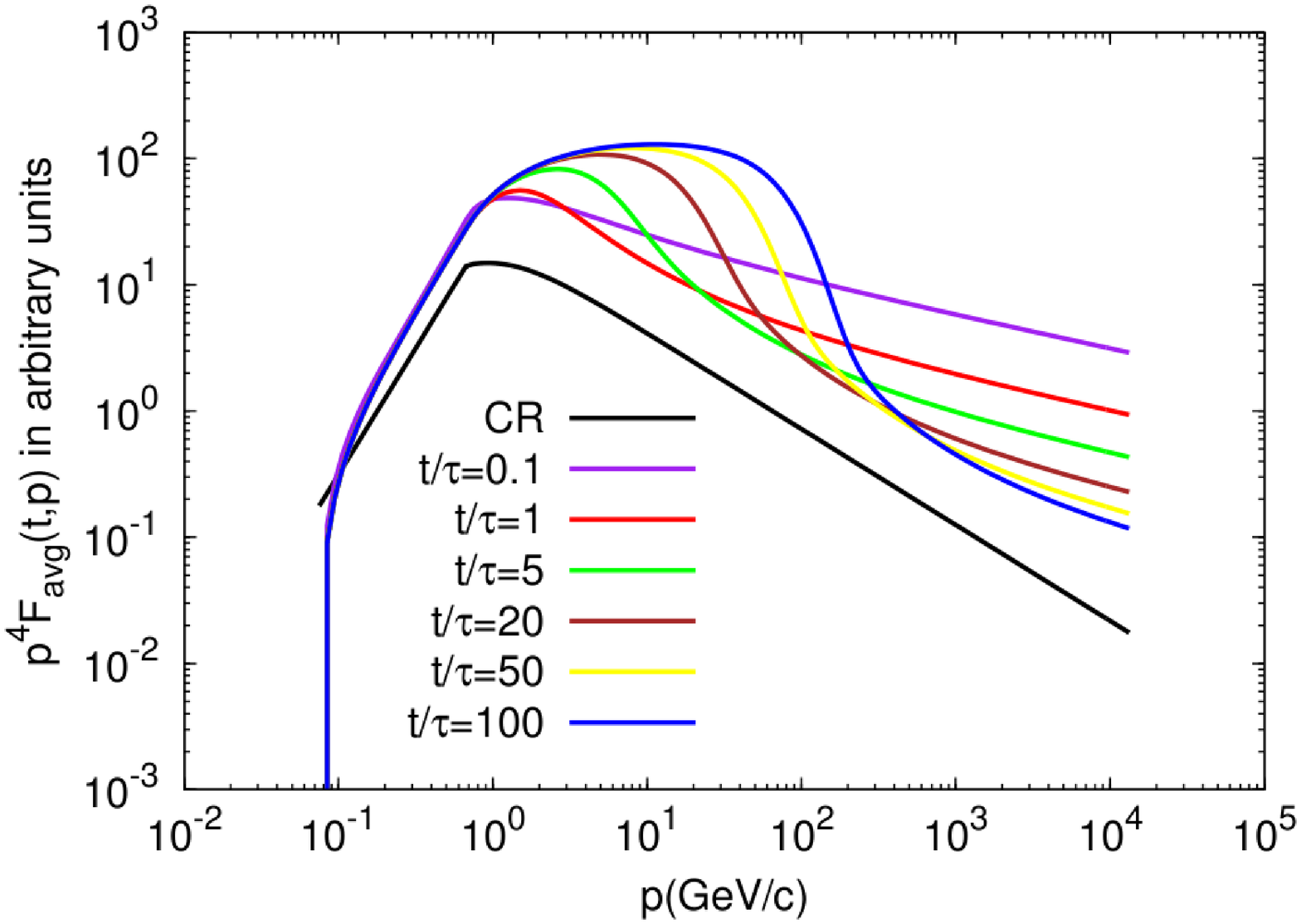} }
 \hfill  
 \subfigure[]    
 {\label{DSA_ED_05}
  \includegraphics[width=0.48\textwidth]{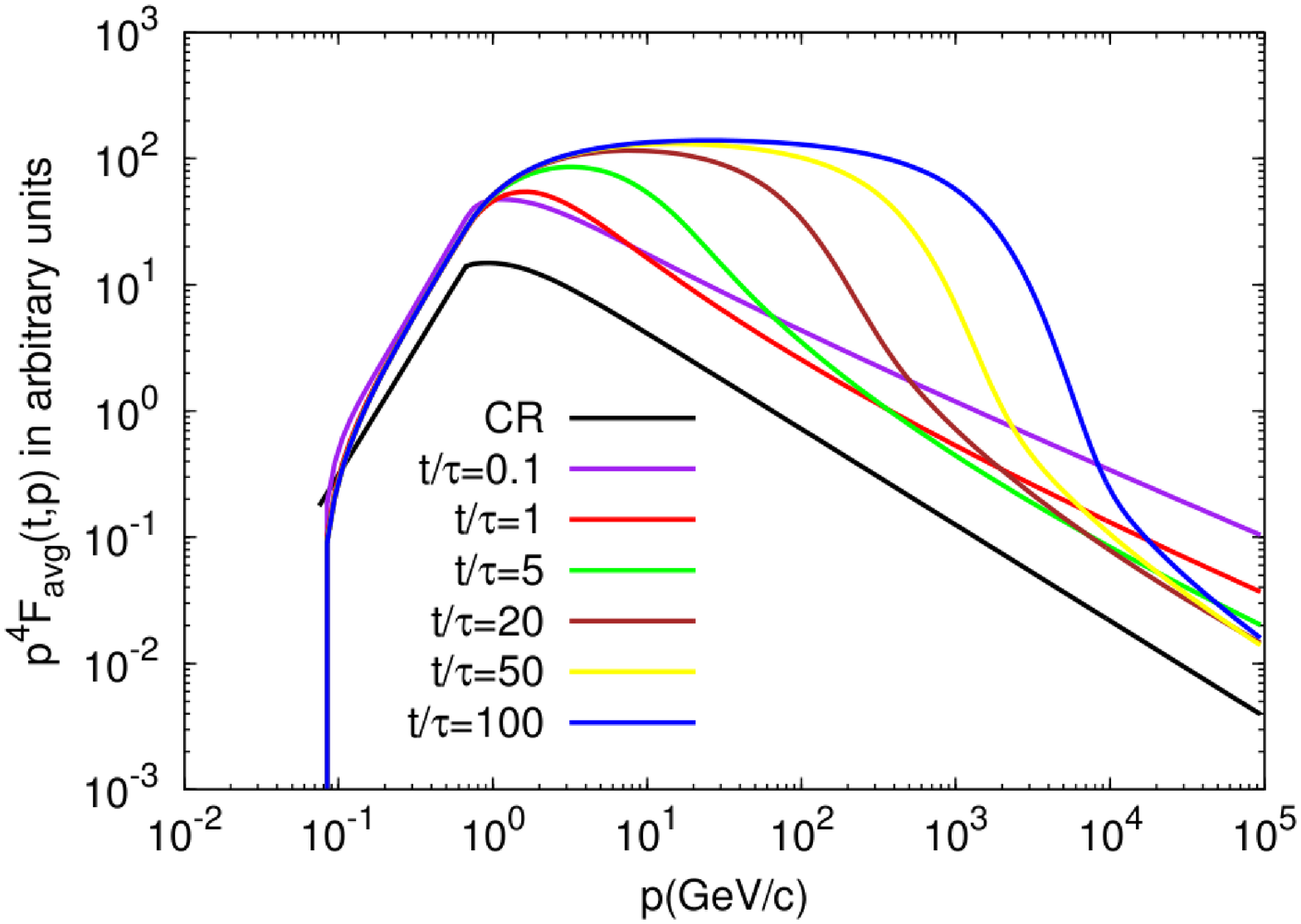}} 
\end{center}
\caption{Time evolution of the spatially averaged downstream particle momentum spectrum. The cosmic ray spectrum denoted by CR has arbitrary scaling. (a) Bohm-like diffusion with diffusion coefficient $\kappa \propto p$; (b) diffusion coefficient $\kappa \propto p^{0.5}$.}
\end{figure}

We  used Talbot's method to calculate the spatially averaged  downstream particle spectrum $F_{avg}(t,p)$ for  energy dependent diffusion.   Here, we are particularly interested in two cases: Bohm-like diffusion with $\sigma=1$, and  $\sigma=0.5$,  which is consistent with  observations of Galactic CR \citep[e.g.,][]{Berezinskii90}.   The resulting particle spectra for the two cases are shown in Figs.\ \ref{DSA_ED_1} and \ref{DSA_ED_05}, respectively, for various time ratios at $p=\rm 1~GeV/c$.  In both spectra there is  a critical momentum below which the spectrum reaches the steady state solution as in the energy independent diffusion case, while above the critical momentum, the particle spectrum  gradually hardens by a  $\sigma/2$ power of momentum compared to the input CR spectrum. This is because above the critical momentum, $\theta(p)=t/\tau(p)$ becomes much smaller than 1 and the particle motion is dominated by the diffusion process. When diffusion dominates, the length scale for particle motion $L_{diff}\sim \sqrt{6\kappa t}\propto p^{\sigma/2}$. As a result, over a certain time interval $t$, high energy particles moving into the downstream region can trace back to a region further away in the preshock medium, which   hardens the spectrum by a $\sigma/2$ power.

In \cite{Tang14},  we found that the observed $\gamma$-ray emission from middle aged SNRs like IC 443, W28 and W51C implies an accelerated particle spectrum that is similar in shape to the pre-existing CR spectrum at high energy. In the above discussion, we have shown that the time dependent DSA solution in the test particle limit naturally produces an accelerated particle spectrum similar to the input CR spectrum at high energy when $\theta$ is not large. The resulting particle spectrum follows the steady state solution at low energy, but at high energy the particle spectrum is determined by both the input CR spectrum and the possible energy dependent diffusion. In the energy independent diffusion case, the spectrum simply recovers the steep power law shape of the input CR spectrum when approaching high energy, while for energy dependent diffusion the power law shape of the input CR spectrum gradually hardens by a $\sigma/2$ power at high energy.

\section{PION-DECAY EMISSION}

Assuming the $\gamma$-ray emission from those middle aged SNRs interacting with molecular clouds  has a hadronic origin \citep[e.g.,][]{Ackermann13}, here we estimate the $\pi^0$-decay emission from IC 443 and W44 based on our time dependent DSA solution and then compare the emission with observations. We take the model in \cite{Tang14} for molecular clump interaction, which is simplified from the following picture in \cite{chevalier99}:
the remnant becomes radiative in the interclump medium of the molecular cloud, forming a cool shell, and the shell collides with dense molecular clumps, producing a layer of shocked shell (layer 1) and a layer of shocked molecular clump (layer 2). The radiative shell, layer 1 and layer 2 are all potential sources of $\gamma$-ray emission.
According to the calculations in \cite{Tang14}, emission from layer 2 is much smaller than the other two components, so in the following discussion we only model the $\pi^0$-decay emission from the radiative shell and layer 1.

Considering an emission region of volume $V$ with uniformly distributed ambient protons of number density $n_a$ and accelerated CR protons of number density $n_{acc}$, the resulting $\pi^0$-decay luminosity from the system is
\begin{equation}
L_{\pi^0}(E_{\gamma},t)
=\chi E_{\gamma}c\int^\infty_{E_{p,thresh}}dE_{p} \beta_{p}\frac{d\sigma (E_{\gamma},E_p)}{dE_{\gamma}}n_{a}(t) n_{acc}(t,E_{p})V,
\label{emission}
\end{equation}
where $E_p$ is the CR proton energy, $E_{\gamma}$ is the emitted photon energy, $t$ is the age of the system, $d\sigma(E_{\gamma},E_p)/dE_{\gamma}$ is the $\pi^0$-decay cross section, and $\chi$ is the scaling factor for helium and heavy nuclei which is taken to be 1.8 \citep{mori09}. For this estimate,  we ignore the dynamic evolution of the remnant and the accompanying particle loss through escape and energy loss through radiative cooling. CR particles are  accelerated through both DSA and adiabatic compression. For seed particles with number density $n_{seed}$, we define the number density of CR that undergoes DSA for a time interval of $t$ as $n_{DSA}(t,p,n_{seed})=4\pi p^2F_{avg}(t,p)$.   Then the accelerated CR spectrum with both DSA and adiabatic compression becomes 
\begin{equation}
n_{acc}(t,E_{p})=\frac{n_{acc}(t,p)}{v(p)}=\frac{(\lambda/4)^{2/3}n_{DSA}(t,(\lambda/4)^{-1/3}p,n_{seed})}{v(p)},
\end{equation}
where $v(p)$ is the particle velocity and $\lambda$ is the total compression ratio for the emission region.

For the radiative shell, $t$ equals the age of the remnant  $t_{age}$,  $n_a(t_{age})$ equals the shell density $n_s(t_{age})$, the seed particles are the pre-existing CRs which are taken to be the same as $n_{GCR}$ in \cite{Tang14}, and the emission volume is $V\approx4\pi R(t_{age})^3$ $(1-\eta)/3\lambda_s$, where $R(t_{age})$ is the remnant radius, $\lambda_s=n_s/n_0$ is the shell compression ratio, and $\eta$ is the volume filling factor for molecular clump interaction.  The resulting accelerated CR number density in the shell then becomes 
\begin{equation}
n_{acc,\, s}(t_{age},E_{p})=\frac{n_{acc,\, s}(t_{age},p)}{v(p)}= \frac{ (\lambda_s/4)^{2/3} n_{DSA,\, s}(t_{age},(\lambda_s/4)^{-1/3}p,n_{GCR})}{v(p)}.
\end{equation} 

For layer 1, $t=t_{MC}$ which is the time since molecular clump interaction started and is taken to be the time that layer 1 is about to break out of the shell \cite[see details in][]{Tang14}, $n_a(t_{MC})$ becomes the density of layer 1, $n_{l_1}(t_{MC})$, and the emission volume $V\approx4\pi R(t_{age})^3\eta /3\lambda_s\lambda_{l_1}$, where $\lambda_{l_1}=n_{l_1}/n_s$ is the layer 1 compression ratio.   The number density of seed particles now becomes $n_{acc,\, s}$ and the resulting accelerated CR number density in layer 1 is
\begin{equation}
n_{acc,\, l_1}(t_{MC},E_{p})=\frac{n_{acc,\, l_1}(t_{MC},p)}{v(p)}= \frac{(\lambda_{l_1}/4)^{2/3}n_{DSA,\, l_1}(t_{MC},(\lambda_{l_1}/4)^{-1/3}p,n_{acc,\, s})}{v(p)}.
\end{equation} 

Pre-existing CR in the ambient medium undergo two periods of DSA in reaching layer 1, so the resulting accelerated CR spectrum in layer 1 is determined by two time ratios $\theta_f$ and $\theta_{l_1}$, which correspond to the SNR forward shock and the layer 1 shock, respectively.  Due to the two  DSA episodes, the resulting CR spectrum for energy dependent diffusion is hardened by one $\sigma$ power instead of $\sigma/2$ power at high energy.   The time dependent DSA solution we derived here is under the assumption that there are seed particles uniformly distributed in the preshock medium extending to infinity. This is a good assumption for the SNR forward shock but, for the layer 1 shock sweeping up the radiative shell material,  it breaks down especially when layer 1 is about to break out of the shell.   Here we use the time dependent solution for both the SNR forward shock and the layer 1 shock for simplicity. We expect the hardening at high energy will be less significant if the seed particles are only distributed in a limited size region of preshock medium.

Following the procedure in \cite{Kamae06},  we calculate the $\pi^0$-decay emission from IC 443 and W44, and then use the results as examples to show that simple estimates based on time dependent DSA and $\pi^0$-decay emission can reproduce the $\gamma$-ray emission with the observed overall spectral shape.   Combining the observational data and the molecular interaction model in \cite{Tang14} we can obtain the parameters for the SNR and the molecular interaction region, leaving only three variables  in our spectrum fitting:  the remnant forward shock time ratio $\theta_f$, the layer 1 shock time ratio $\theta_{l_1}$, and the volume filling factor $\eta$.   The time ratios $\theta_f$ and $\theta_{l_1}$ obtained through spectrum fitting can be further used to estimate the diffusion coefficient of CR particles around the remnant and the molecular interaction region.  By definition $\theta_f=t_{age}U_f^2/4\kappa_0$, where $U_f$ is the forward shock velocity and $\kappa_0$ is the diffusion coefficient in the upstream region of the forward shock, while $\theta_{l_1}=t_{MC}(U_f-U_{l_1})^2/4\kappa_s$, where $U_{l_1}$ is the layer 1 shock velocity and $\kappa_s$ is the diffusion coefficient in the upstream region of the layer 1 shock.   For typical parameters in middle aged SNRs and the molecular interaction region, we find 
\begin{equation}
\theta_f\approx8\left( \frac{t_{age}}{\rm 10^4~yrs}\right)\left( \frac{U_f}{100 \kms} \right)^2\left( \frac{10^{24} \rm~ cm^2~s^{-1}}{\overline{\kappa_0}}\right)\left(\frac{1\rm~ GeV/c}{p} \right)^\sigma
\end{equation}
and
\begin{equation}
\theta_{l_1}\approx0.08\left( \frac{t_{MC}}{\rm 100~yrs}\right)\left( \frac{U_f-U_{l_1}}{100 \kms} \right)^2\left( \frac{10^{24} \rm ~cm^2~s^{-1}}{\overline{\kappa_s}}\right)\left(\frac{1\rm~ GeV/c}{p} \right)^\sigma
\end{equation}
under the assumption that $\kappa=\overline{\kappa}\,\overline{p}^{\sigma}$. $\overline{\kappa_0}$ and $\overline{\kappa_s}$ are unknown parameters depending on the shock environment, especially the surrounding magnetic irregularities, and may be related to each other.
For example, in the special case of Bohm-like diffusion, $\overline{\kappa}=\xi pc^2/3eB$ where $\xi>1$ is the gyro-factor. If we assume $\xi$ is  constant within the SNR and molecular interaction region, then $\overline{\kappa_0}/\overline{\kappa_s}=B_s/B_0$, where $B_0$ is the magnetic field in the ambient medium and $B_s$ is the magnetic field in the radiative shell. As a result, $\theta_f/\theta_{l_1}=t_{age}U_f^2B_0/t_{MC}(U_f-U_{l_1})^2B_s\approx 1$ for both IC 443 and W44 with our parameters.
For energy dependent diffusion with arbitrary power law index $\sigma$, there is no theory for  the ratio $\theta_f/\theta_{l_1}$, so we leave $\theta_f$ and $\theta_{l_1}$ as two independent parameters for our fits. We do require that the ratio $\theta_f/\theta_{l_1}$ fall between the value from Bohm-like diffusion and the value from assuming $\overline\kappa_0=\overline{\kappa_s}$.   
In fitting the spectrum, we allow for three different situations: energy independent diffusion with $\sigma=0$, Bohm-like diffusion with $\sigma=1$, and energy dependent diffusion with $\sigma=0.5$. For Bohm-like diffusion we apply the relation $\theta_f=\theta_{l_1}$, as discussed above, while for the other two cases we require $1\lesssim\theta_f/\theta_{l_1}\lesssim t_{age}U_f^2/t_{MC}(U_f-U_{l_1})^2$.   In the above calculation, we ignore the dynamical evolution of the SNR and assume $U_f(t)=U_{f}(t_{age})$.  This might affect our estimate of $\overline{\kappa_0}$, but our result can at least provide order of magnitude information on the diffusion coefficient because $U_f$ has a weak dependence on time. In our spectral modeling, we have not attempted to obtain a best fit in view of the complex physical situation and model uncertainties, but aim to show the importance of the time dependent DSA solution in improving the fit to the spectrum. Self-consistent models or simulations with time dependent DSA coupled with the dynamical evolution of the SNR are required in the  future for detailed comparisons with observations. 

\begin{figure}[htb]   
 \begin{center}
\subfigure[]
 {\includegraphics[width=0.48\textwidth]{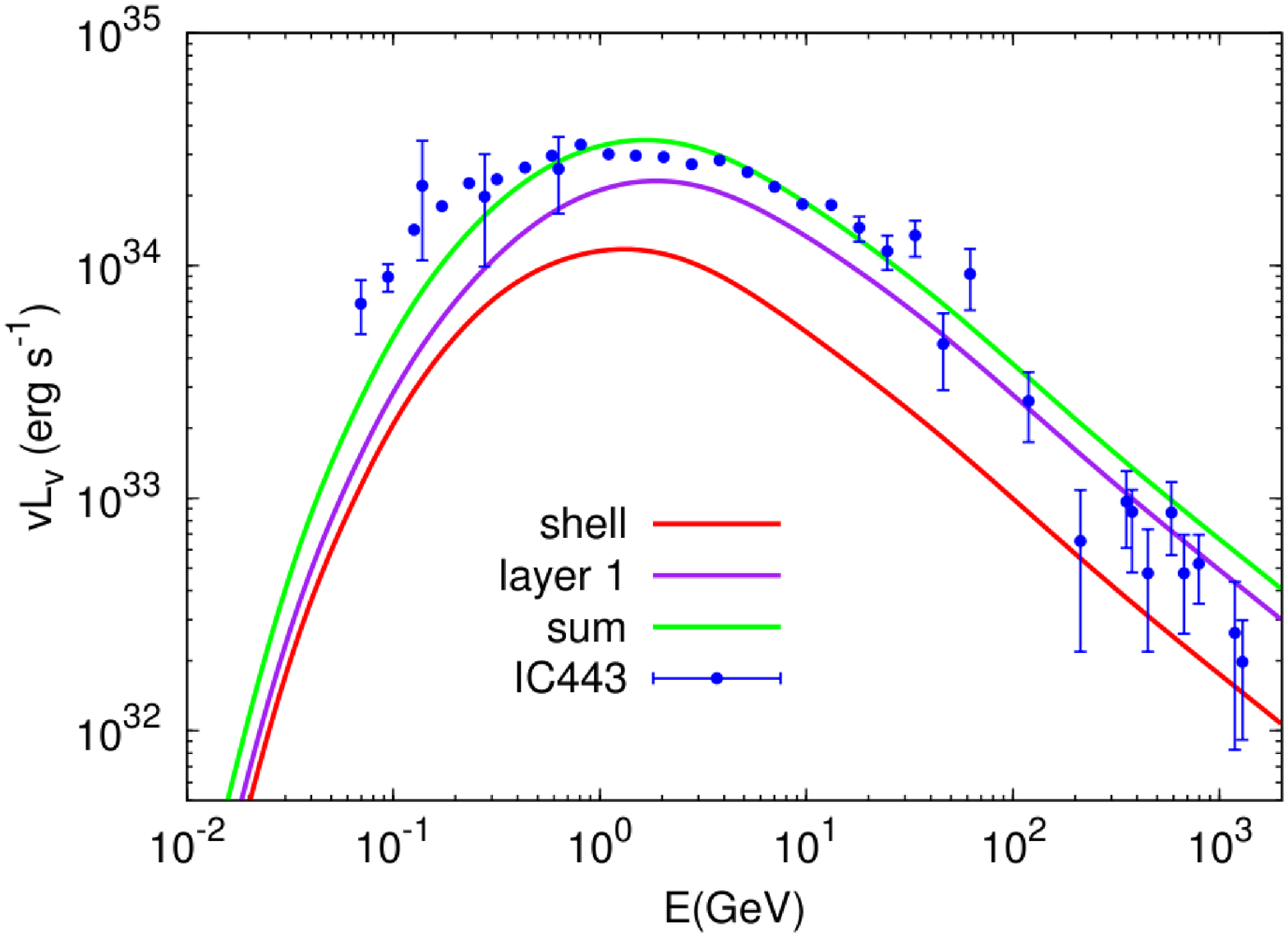} 
    \label{DSA_IC443}}
    \hfill
\subfigure[]
{ \includegraphics[width=0.48\textwidth]{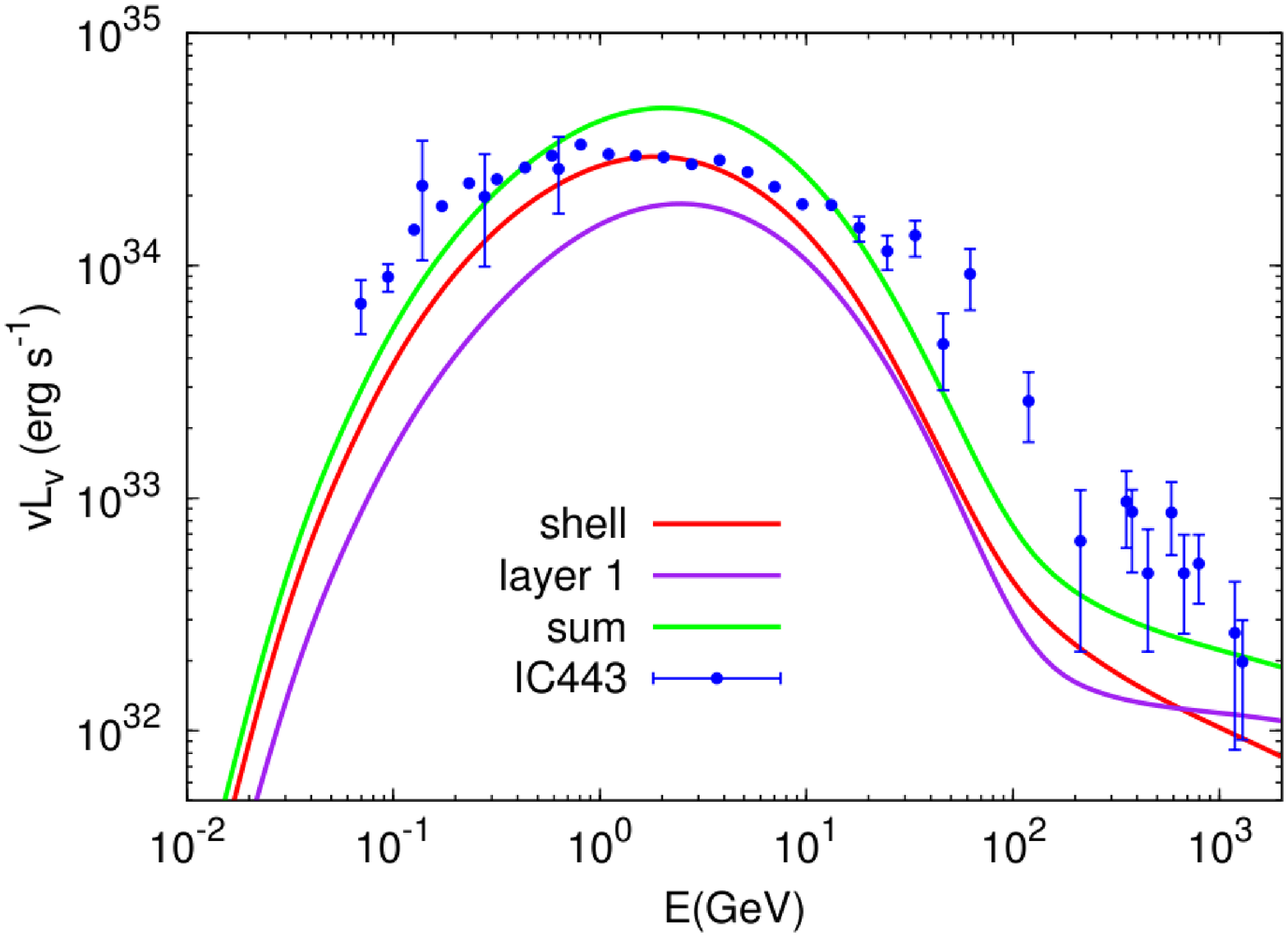}  
    \label{IC443_p1}}
\\ 
 \end{center}
 \subfigure[] 
 {\includegraphics[width=0.48\textwidth]{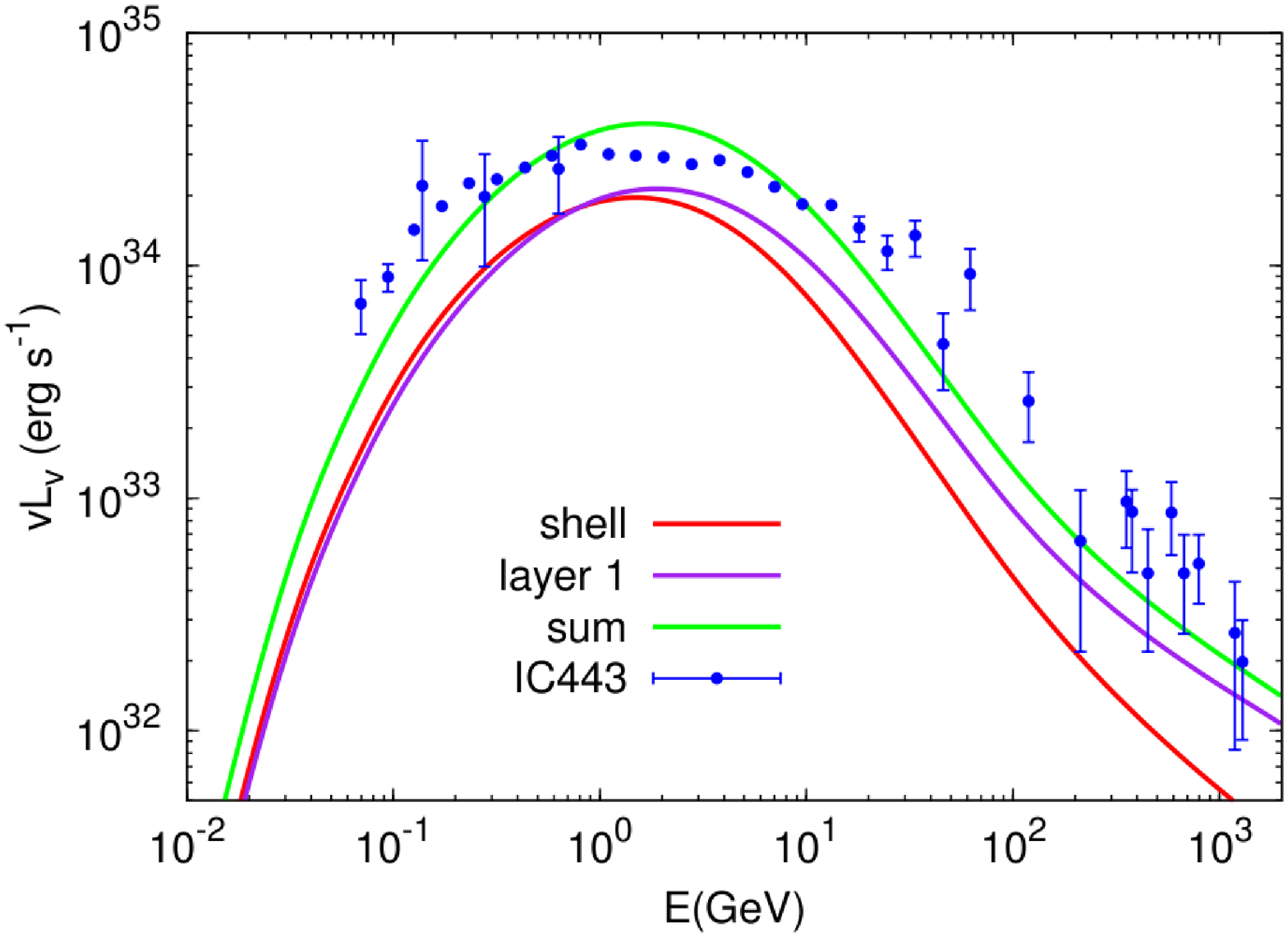} 
    \label{IC443_p05}}

 \caption{$\pi^0$-decay emission from IC 443 for different energy dependence of diffusion coefficient. Shell is the radiative shell of remnant, layer 1 is the shocked shell, and sum is the sum of the 2 components. (a) Energy independent diffusion with $\theta_f=2\theta_{l_1}=2$ at $p=1{\rm ~GeV}/c$ and $\eta=0.2$ compared to observations; (b) energy dependent diffusion with $\kappa \propto p$, $\theta_f=\theta_{l_1}=40$ at $p=1{\rm ~GeV}/c$ and $\eta=0.06$; (c) energy dependent diffusion with $\kappa \propto p^{0.5}$, $\theta_f=16\theta_{l_1}=8$ at $p=1{\rm ~GeV}/c$ and $\eta=0.15$. The data points are taken from the same references as in \cite{Tang14}.}
 \label{IC443}
 \end{figure}

For IC 443,  we use parameters from Table 1 in \cite{Tang14} for the remnant and molecular interaction region. In the energy independent diffusion case,  we show an example fit with $\theta_f=2\theta_{l_1}=2$ at $p=1{\rm ~GeV}/c$ and a filling factor $\eta=0.2$ (Fig.\ \ref{DSA_IC443}). The resulting spectrum is similar to the pure adiabatic compression case in \cite{Tang14}, since DSA with energy independent diffusion can reproduce the input CR spectrum at small $\theta$. In the energy independent diffusion case, the spectral shape is mainly determined by $\theta_f$ and is not very sensitive to $\theta_{l_1}$, which is coupled with $\eta$. For Bohm-like diffusion with $\sigma=1$ and assuming $\theta_f\approx\theta_{l_1}$, we can roughly fit the observed $\gamma$-ray emission with $\theta_f=\theta_{l_1}=40$ at $p=1{\rm ~GeV}/c$ and $\eta=0.06$ (Fig\ \ref{IC443_p1}). The spectral hardening at high energy produces too flat a spectrum compared to observations, which implies that Bohm-like diffusion is probably not a good assumption for these middle aged SNRs in the context of our model.  For $\sigma=0.5$, an example fit with $\theta_f=16\theta_{l_1}=8$ at $p=1{\rm ~GeV}/c$ and $\eta=0.15$ is presented in Fig \ref{IC443_p05}. The $\pi^0$-decay emission due to energy dependent diffusion is characterized by a break at high energy compared to the energy independent diffusion case. For IC 443 the break is likely to be around 100 GeV according to our fit. The significance of the break is simply determined by $\sigma$. It is clear that energy dependent diffusion with $\sigma=0.5$ fits the observations better than Bohm-like diffusion. This is comparable to the $\sigma$ value inferred for Galactic CR \citep{Berezinskii90}.   With the values of $\theta_f$ and $\theta_{l_1}$ obtained above, we can estimate the diffusion coefficient of CR particles around the SNR and the molecular interaction region. We only discuss the CR diffusion coefficient around the remnant forward shock because $t_{MC}$ for the molecular interaction is uncertain, although we assume $t_{MC}$ equals the break out time to simplify the calculation. The calculated values of $\overline{\kappa_0}$ for $\sigma=0$, $\sigma=0.5$, and $\sigma=1$ at $p=\rm 1~GeV/c$ are $9\times 10^{24}\rm ~cm^2~s^{-1}$, $2\times 10^{24}\rm ~cm^2~s^{-1}$, and $ 4\times 10^{23}\rm ~cm^2~s^{-1}$ respectively, which are much smaller than the CR diffusion coefficient at $p=\rm 1GeV/c$ in the ISM, $\sim 3\times 10^{27} \rm ~cm^2~s^{-1}$ \citep{Berezinskii90}, but are closer to the Bohm limit at $p=\rm 1~GeV/c$, $\sim 7\times 10^{21}\rm ~cm^2~s^{-1}$. According to the discussion in Section 2, for the SNR forward shock our simplified model parameter $\tau=\tau_1=\tau_2\approx \hat{\tau_1}$ reflects the value of the spatially averaged diffusion coefficient in the upstream region. The CR diffusion coefficient close to the shock front should be lower than the estimated $\overline{\kappa_0}$ above and close to the Bohm limit.

W44 (G34.7 - 0.4) is a mixed morphology SNR with centrally filled X-ray emission and shell-like radio emission. The distance to the remnant is estimated to be $\sim 3$ kpc based on both HI 21cm absorption measurements \citep{Caswell75} and molecular observations \citep{Castelletti07}. \cite{Wolszczan91} discovered a 267 msec pulsar, PSR 1853 + 01, in the southern part of W44 well within its radio shell. The pulsar has a spin down age $\sim 2\times 10^4 $ years and a dispersion measure distance  consistent with the remnant distance, which implies the pulsar is likely to be associated with the W44. The remnant is elongated with a size of $11\times 15$ pc at a distance of 3 kpc, so we take 13  pc for the remnant radius as in \cite{chevalier99}. The forward shock velocity is taken to be $150 \kms$ since \cite{Koo95} found an expanding HI shell moving at velocity of $150  \kms$, which may be the expanding cool shell formed in the radiative phase. Millimeter wavelength observations of  CO and CS lines indicate a molecular shock velocity of $20-30 \kms$ \citep{Reach05}, so we take a molecular clump shock velocity $U_c$ of $30 \kms$ in our calculation. The preshock magnetic field is taken to be 6 $\mu$G, similar to \cite{Tang14}. W44 has a $\gamma$-ray luminosity about one order of magnitude higher than IC 443.  In order to obtain such a high $\gamma$-ray luminosity we require a larger SN explosion energy, $\sim 3 \times 10^{51}$ erg. The other parameters for the remnant  and the molecular interaction region can be obtained from the radiative SNR model in \cite{Cioffi88} and the molecular clump interaction model in \cite{Tang14}, respectively.  The parameters  we use for W44 are listed in Table \ref{W44MC}. 

\begin{figure}[htb]
 \begin{center}
 \includegraphics[width=\textwidth]{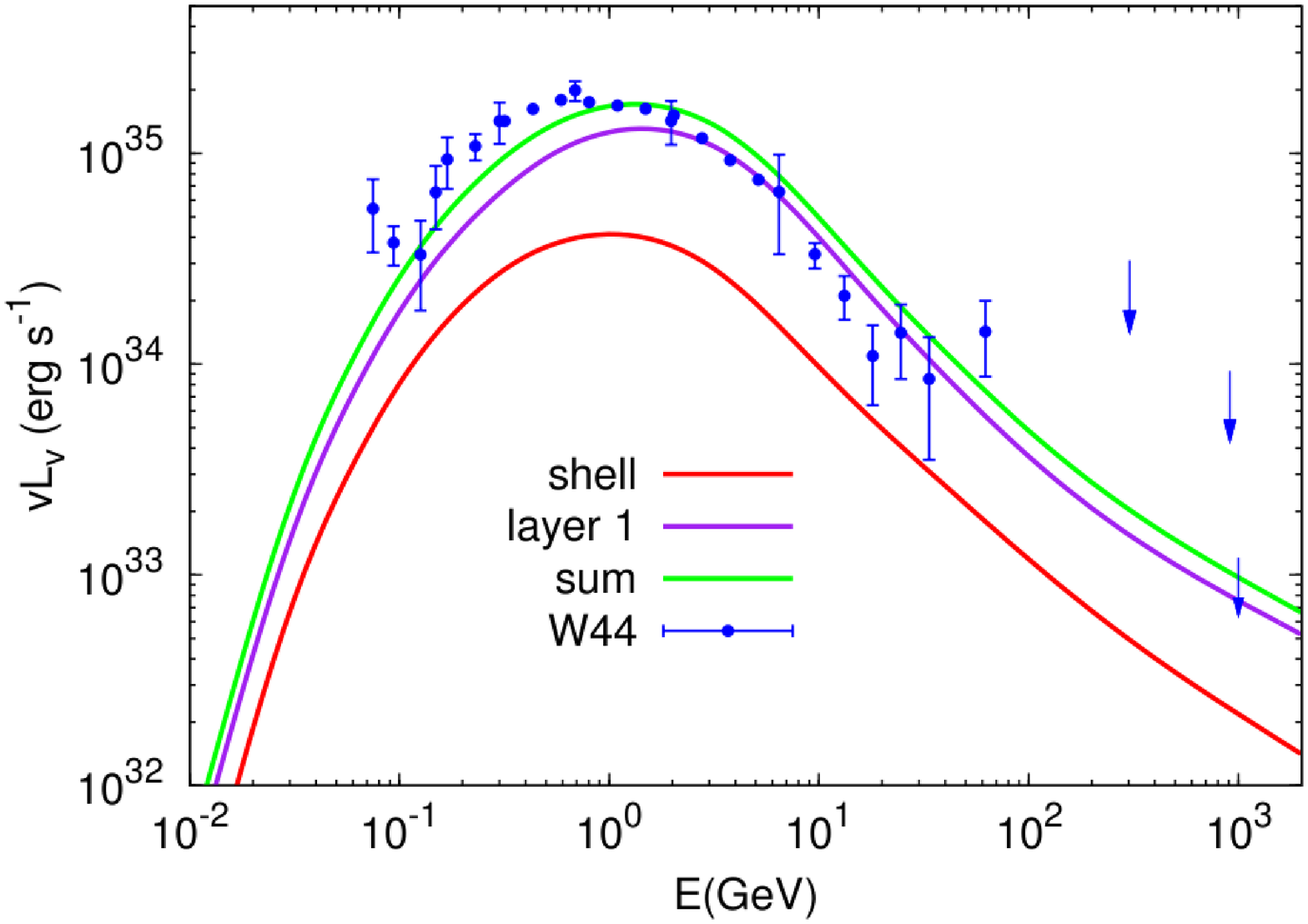} 
  \caption{Like Fig.\ \ref{IC443} but for the W44 remnant and models with $\kappa \propto p^{0.5}$, $\theta_f=\theta_{l_1}=3$ at $p=1{\rm ~GeV}/c$ and $\eta=0.3$. The data points are taken from \cite{Ackermann13,Buckley98,Aharonian02,ong09}} 
    \label{W44_p05}
 \end{center}
 \end{figure}

W44 has a steeper $\gamma$-ray spectrum than IC 443 in the GeV range \citep{Ackermann13}, while in the TeV range there are only upper limits so far \citep{Buckley98,Aharonian02,ong09}.
The steep spectrum  above 1 GeV makes it difficult to fit the W44 data with an energy independent diffusion model as it reproduces the pre-existing CR spectrum at high energy which has a shallower shape.  We focus our attention on the energy dependent diffusion cases.  Bohm-like diffusion with $\sigma=1$ produces too flat a spectrum at high energy which is also disfavored by the data, so for W44 we only show the result for $\sigma=0.5$. An example fit with $\theta_f=3\theta_{l_1}=3$ at $p=1{\rm ~GeV}/c$ and $\eta=0.3$ is shown in Fig.\ \ref{W44_p05}.    The corresponding $\overline{\kappa_0}\approx 2\times 10^{25} \rm ~cm^2s^{-1}$  at $p=1{\rm ~GeV}/c$ is about one order of magnitude larger than that in IC 443 and is about 3000 times larger than the Bohm diffusion coefficient at $p=1{\rm ~GeV}/c$.  In the fit for W44, our TeV spectrum is  close to the upper limit provided by VERITAS \citep{ong09}.  However, there are  factors that could reduce the emission in the TeV range.   For the layer 1 shock the pre-existing CR are only distributed in a limited  region, which could soften the spectrum at high energy. A smaller $\sigma$ could also soften the spectrum at high energy.

\section{DISCUSSION}

We have obtained a time dependent DSA solution in the test particle limit for a planar parallel shock with pre-existing CRs in the preshock region. By combining the time dependent DSA solution derived here and the molecular clump interaction model in \cite{Tang14}, we can produce $\pi^0$-decay emission that compares well to observations.   The derived time ratio $\theta_f$ can be further used to estimate the diffusion coefficient of CR particles around the SNR, but the estimated diffusion coefficient should be considered as a spatially averaged value and be taken as an upper limit for the diffusion coefficient near the shock front.  We discussed three situations for our time dependent DSA solution:  energy independent diffusion,  Bohm-like diffusion with energy index $\sigma=1$, and energy dependent diffusion with $\sigma=0.5$.  For both IC 443 and W44, the best fit is with  energy dependent diffusion with $\sigma=0.5$, which is roughly consistent with Galactic CR observations. The resulting time dependent DSA spectrum is characterized by a critical energy below which the spectrum reaches the steady state solution while above it the spectrum recovers the steep power law shape of the pre-existing CR spectrum with possible hardening due to energy dependent diffusion. Based on the above spectral shape we expect the $\gamma$-ray emission from these middle aged SNRs interacting with molecular clouds to show a spectral hardening in the TeV range which might be detectable by future advanced instruments. If observed, the hardening could be used to derive information about CR diffusion around the SNR shock.  

\cite{Malkov11} have shown that the steep spectrum of W44 can be explained if accelerated particles can escape from the shock region due to the ion neutral damping mechanism, which steepens the spectrum by exactly one power.  Under the assumption of the test particle limit, the high shock velocity, $\ga 120 \kms$, in W44 is inconsistent with the weakly ionized preshock medium required for ion neutral damping \citep{Hollenbach89}. But if  non-linear effects are strong,  efficient CR acceleration and escape could modify the shock structure and allow ion neutral damping in W44 \citep{Bykov14}. A self-consistent model with DSA coupled to the SNR evolution is needed in the future to fully understand the role of ion neutral damping in W44.

Here, we did not take escape of CR particles into account because it may not be important for the energy range of interest. There have been simulations  using CR escape to explain the $\gamma$-ray emission from the middle aged SNRs discussed here \citep[e.g.,][]{Ohira11}, but these models require that accelerated CR particles with energies down to $\sim1~\rm GeV$ escape from the remnant and illuminate the nearby dense clump.  This assumption needs more detailed investigation. Here we use Bohm diffusion as an example, because the diffusion coefficients we estimated are close to the Bohm diffusion limit and there is also observational evidence  indicating possible Bohm diffusion in young SNR \citep[e.g.,][]{Uchiyama07}. Following the discussion in \cite{Ohira11}, the critical momentum for CR particles that can escape the remnant satisfies  $p_{esc}=\kappa D^{-1}_0R_{sh}u_{sh}$ (eq. $(17)$ in \cite{Ohira11}; see the definitions there for the parameters in the formula).  Assuming that SNRs are the CR accelerators up to the energy of CR knee $\sim 10^{15} \rm ~eV$  i.e. $p_{esc}(t_{Sedov})\sim10^{15}\rm ~eV/c$, where $t_{Sedov}$ is the transition time from the free expansion phase to the Sedov-Taylor phase, then the critical momentum for escaping CR particles at $t_{age}$ now becomes $p_{esc}(t_{age})=p_{esc}(t_{Sedov})R_{sh}(t_{age})u_{sh}(t_{age})D_0(t_{Sedov})/D_0(t_{age})R_{sh}(t_{Sedov})u_{sh}(t_{Sedov})$. The escape models developed so far  focus on the Sedov-Taylor phase of the SNR in which $R_{sh}\propto t^{2/5}$ and $u_{sh}\propto t^{-3/5}$. In such a situation, 
\begin{equation}
p_{esc}(t_{age})=p_{esc}(t_{Sedov})\left(\frac{t_{age}}{t_{Sedov}}\right)^{-1/5}\frac{B_{t_{age}}}{B_{t_{Sedov}}}.
\end{equation}
As a result, for the evolution of a middle aged SNR from $t_{Sedov}\sim 100$ yrs to $t_{age}\sim  10^4$ yrs and a magnetic field amplification factor of $B_{t_{Sedov}}/B_{t_{age}}\sim 100$, the critical momentum of escaping CR particles right now is $t_{age}\approx 4\times 10^{12} ~\rm eV/c$, above the energy range discussed here.  The escaping CR particles which reach the nearby dense clumps and illuminate them would have even higher energy.  Obtaining $p_{esc}(t_{age})\sim \rm 1~GeV/c$ requires extreme conditions for  parameters like the magnetic field amplification factor, or the diffusion coefficient of CR particles must have a weak dependence on particle momentum and relatively strong dependence on magnetic field, which is not clear from observations.

In our model we assume a parallel shock for simplicity, but in reality the magnetic field in the ambient medium is likely to be randomly distributed while the molecular shock is likely to be a perpendicular shock due to a magnetically  supported shell. For an oblique shock with angle $\phi$ between the magnetic field direction and the shock normal, the diffusion coefficient $\kappa=\kappa_{||}\cos^2\phi + \kappa_{\perp}\sin^2\phi$  where $\kappa_{||}$ is the diffusion coefficient along the magnetic field lines and $\kappa_{\perp}$ is the diffusion coefficient across the field lines   \citep[e.g.,][]{Reynolds98}. 
In general, if we take the obliquity of the shock into account it would affect our estimate of the CR diffusion coefficient depending on the angle $\phi$ and the relation between $\kappa_{||}$ and $\kappa_{\perp}$, but it does not affect the time ratios $\theta_f$ and $\theta_{l_1}$ derived in the fits to spectra.

In our molecular clump interaction model we only consider the situation that layer 1 has not broken out of the radiative shell.  In reality, layer 1 could break out of the    
radiative shell after a sufficient time of interaction.
In that case the emission from layer 2 might become dominant. Unlike the shocked shell matter in layer 1, the shocked clump matter accumulated in layer 2 only undergoes one episode of DSA, which produces a spectrum with less hardening, by $\sigma/2$ at high energy compared to layer 1. As a result, when layer 2 dominates the $\gamma$-ray emission, the Bohm-like diffusion case  would produce a steeper spectrum and fit the observations better. \cite{Anderl14} found evidence for non-stationary shocks in W44  with age $\sim 10^3$ yrs 
 through a radiation transfer model of the CO(7-6) and CO(6-5) transitions.
The ages suggest that in W44 layer 1 may already have broken out of the radiative shell.  

In our spectral fits for both IC 443 and W44, the $\gamma$-ray emission from layer 1 is either comparable to or larger than the emission from the shell component. Considering the small filling factor $\eta$ in the fit, emission from layer 1 would have a larger $\gamma$-ray surface brightness than the shell. After projection effects, the shell is expected to show a ring-like or filamentary structure in $\gamma$-rays while the morphology for molecular interaction region could be complex. Instead of interacting with one single large clump, the remnant is likely to be interacting with multiple clumps at the same time.   The $\gamma$-ray morphology of the molecular interaction region is also determined by the angle between the molecular shock normal and the viewing angle direction. If the shock normal is perpendicular to the line of sight, we would expect $\gamma$-ray morphology with a ring or arc-like feature plus some bright spots on the edge of the ring.  If the molecular shock normal is more or less along the line of sight, we might observe a roughly uniform disk-like morphology or center bright morphology with multi-clump interaction. In order to disentangle all the different situations we require more detailed observations of the molecular interaction region. 

Finally, we note that the time dependent DSA model presented here should also be applicable to other interaction models with re-acceleration of pre-existing CR in the preshock medium  
\citep[e.g.,][]{Uchiyama10}. 

\acknowledgements
We thank A. Bykov for comments and the referee for clarification of the paper.
This research was supported in part by NASA {\it Fermi} grant NNX12AE51G.

\appendix
\section*{Appendix}
In order to elucidate our results on the spectral shape, we need to understand the micro-physics of the DSA process. In DSA, particles are bouncing back and forth across and around the shock discontinuity as a result of the magnetic turbulence. Every time a particle comes across the shock discontinuity it receives a mean momentum gain $\Delta p=2(U_1-U_2)p/3v$, where $v$ is the particle velocity, and the mean time taken for a particle to complete one cycle of back and forth motion is $\Delta t=4\left( \kappa_1/U_1 +\kappa_2/U_2\right)/v$ \citep{Drury83}. The corresponding momentum gain rate for a particle undergoing DSA is then 
\begin{equation}
\frac{dp}{dt}=\frac{2\Delta p}{\Delta t}=\frac{U_1-U_2}{3}\, \left( \frac{\kappa_1}{U_1} +\frac{\kappa_2}{U_2}\right)^{-1}  p.
\label{energy_gain}
\end{equation} 

Particles entering the downstream region have a chance to escape the DSA site and move to $+\infty$ in the downstream region due to the advective flow towards the positive $x$ direction. The probability for a particle not returning back to the acceleration site is given by $P=4U_2/v$ \citep{Drury83}. Considering a particle with initial momentum $p_i$, after $n$ cycles of acceleration the particle momentum becomes
\begin{equation}
p_n\sim \prod^n_{k=1}\left[1+\frac{4(U_1-U_2)}{3v_k}\right]p_i,
\end{equation}
leading to
\begin{equation}
 {\rm ln}(p_n/p_i)\sim\frac{4(U_1-U_2)}{3}\sum^n_{k=1}\frac{1}{v_k}.
\end{equation}  
The probability for a particle to stay at the acceleration site after $n$ cycles of back and forth motion is 
\begin{equation}
P_n\sim\prod^n_{k=1} \left( 1-\frac{4U_2}{v_k}\right)
\end{equation}
so that
\begin{eqnarray}
{\rm ln}P_n & \sim & -4U_2\sum^n_{k=1}\frac{1}{v_k}\\
	&=&-\frac{3U_2}{U_1-U_2} {\rm ln}(p_n/p_i).
	\label{escape_probability}
\end{eqnarray}

After a time $t=n\Delta t$, the particle momentum changes from $p_i$ to $p_n$. For constant $U$ and $\kappa$, the energy gain rate $dp/dt\propto p$ which implies that the time taken for a particle to increase its momentum by an arbitrary factor is the same for all  particle momenta.  The energy gain during DSA  simply shifts the input spectrum in the momentum direction by a factor of $p_n/p_i$. Based on the conservation of particle number, the new particle spectrum $R_n(p_n)$ is related to the input CR spectrum $R(p_i)$ by
 \begin{equation}
R_n(p_n)dp_n=R(p_i)P_ndp_i,
\end{equation}
where $P_n$ is the probability for a particle to stay at the DSA site after a time $t=n\Delta t$.
For a strong shock,  $P_n=p_i/p_n$ [equation (\ref{escape_probability})] and, based on our energy gain rate, we have $dp_n/dp_i=p_n/p_i$.   After some calculation we obtain $R_n(p_n)p_n^2=R(p_i)p_i^2$. As the particle number density $R(p)$ and the phase space density $f(p)$ are related by $R(p)=4\pi f(p)p^2$, we  obtain the relation 
$f_n(p_n)p_n^4=f(p_i)p_i^4$ for the downstream particle spectrum at the shock front, which indicates that in the $\log (f(p)p^4)-\log (p)$ plane, the whole DSA process works like a horizontal shift of the function $f(p)p^4$.    The amount of shift is determined by
\begin{equation}
{\rm ln}\left( \frac{p_n}{p_i}\right)=\frac{U_1-U_2}{3}\, \left( \frac{\kappa_1}{U_1} +\frac{\kappa_2}{U_2}\right)^{-1}  t,
\label{shift}
\end{equation}
so that $p_n/p_i$ depends on time $t$ exponentially. As a result, the accumulated CR particle spectrum at the shock front after time $t$ is determined by the sum of the input CR spectrum shifted by various amounts along the $\log(p)$ axis due to various injection times in the $\log(f(p)p^4)-\log(p)$ plane. Because of the exponential dependence on time $t$, in the $\log(f(p)p^4)-\log(p)$ plane all the shifted spectra have the same weight for the sum. 

Based on the shape of the input CR spectrum which follows roughly a broken power law, the accumulated downstream particle spectrum at the shock front would have three parts according to above discussion. The low energy and high energy parts of the accumulated particle spectrum maintain the two power law shape of the input CR spectrum  because all the shifted spectra share the same power law index. At intermediate energies, the accumulated particle spectrum shows a plateau which is due to the break in the input CR spectrum. The plateau starts at the break momentum $p_b$ of the input CR spectrum and ends at the momentum $p_t$, which is determined by equation (\ref{shift}). $p_t$ serves as a critical momentum for the accumulated downstream particle spectrum at the shock front; below $p_t$ the resulting spectrum follows the steady state DSA solution while above $p_t$ the spectrum recovers the steep power law shape of the input CR spectrum at high energy. Our discussion here  only provides the overall shape of the accumulated particle spectrum roughly as all the calculations are based on the mean acceleration time and energy gain.

\clearpage
\begin{table}[ht]
\centering
\caption{Basic parameters for the W44 model}
\begin{tabular}{lr}
\hline\hline
SNR dynamics&\\
\hline
Explosion energy, $E$ & $3 \times 10^{51}$ erg\\
Age, $t_{age}$ &27 kyr\\
SNR radius, $R$ &13 pc\\
Remnant forward shock velocity, $U_{f}$ &150 km/s\\
Shock compression ratio, $\Omega_s$ &4\\
\hline\hline
Molecular clump and interclump medium(ICM) &\\
\hline
Preshock ICM density, $n_0$ & 10.3 cm$^{-3}$\\
Magnetic field in ICM, $B_0$ &6 $\mu$G\\
Molecular clump density, $n_c$ &$1.4 \times 10^4$ cm$^{-3}$\\
Magnetic field in molecular clump, $B_c$ &71 $\mu$G \\
\hline\hline
Radiative shell and molecular clump interaction region &\\
\hline
Discontinuity velocity of clump shock, $U_c$  &30 km/s\\
Molecular clump interaction break out time, $t_{\rm MC}$  &0.3 kyr\\
Density in the radiative shell at $t_{age}$,  $n_s$ & $6.6\times 10^2$ cm$^{-3}$\\
Magnetic field in the radiative shell at $t_{age}$,  $B_{ts}$ &$3.7\times 10^2$ $\mu$G\\
Density in layer 1,  $n_{l_1}$ &$4.9\times 10^3$ cm$^{-3}$\\
Magnetic field in layer 1,  $B_{t1}$ & $2.7\times 10^3$ $\mu$G\\
Layer 1 velocity,  $U_{l_1}$ &11 km/s\\
Shock compression ratio for layer 1, $\Omega_{l_1}$ &3.5\\
Density in layer 2,  $n_{l_2}$ &$6.4 \times 10^5$ cm$^{-3}$\\
Magnetic field in layer 2,  $B_{t2}$ &$2.7\times 10^3$ $\mu$G\\
Layer 2 velocity,  $U_{l_2}$ &31 km/s\\
\hline\label{W44MC}
\end{tabular} 
\end{table}

\begin{figure}[htb]     
 \begin{center}
 \includegraphics[width=\textwidth]{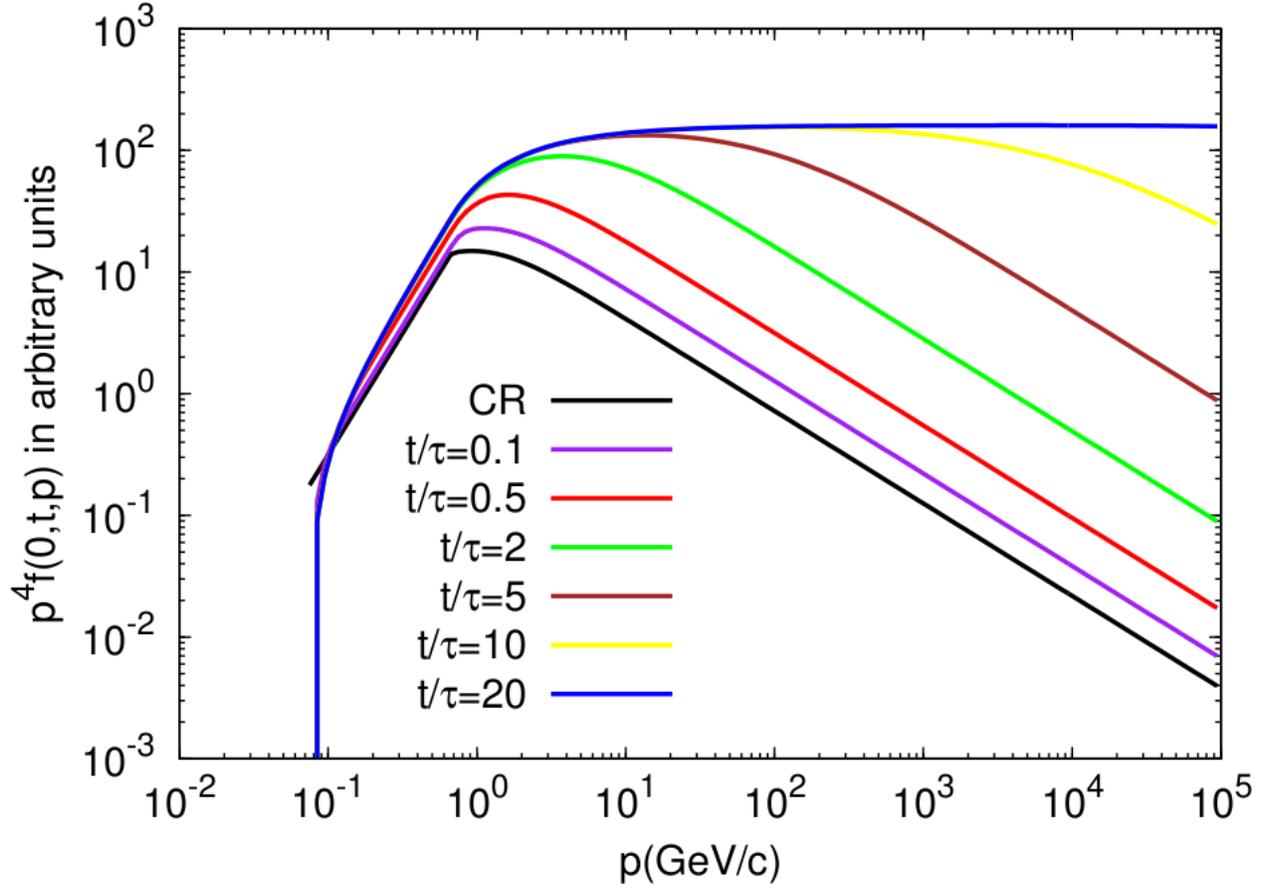} 
    \caption{Time evolution of the particle momentum spectrum at the shock front for various time ratios.  CR denotes the ambient cosmic ray spectrum. The diffusion coefficient is taken to be independent of energy.}
\label{DSA_sf} 
 \end{center}
 \end{figure}

\end{document}